\documentclass[twocolumn,showpacs,prl,preprintnumbers,amsmath,amssymb]{revtex4}
\usepackage{bm}
\usepackage{graphicx}
\usepackage{color}
\begin{document}
\title{InN/In nanocomposites: Plasmonic effects and a hidden optical gap
    }
\author{T.~V.~Shubina$^1$} 
\author{V.~A.~Kosobukin$^1$}
\author{T.~A.~Komissarova$^1$}
\author{V.~N.~Jmerik$^1$}
\author{P.~S.~Kopev$^1$}
\author{S.~V.~Ivanov$^1$}
\author{A.~Vasson$^2$}
\author{J.~Leymarie$^2$}
\author{N. A. Gippius$^2$}
\author{T.~Araki$^3$}
\author{T.~Akagi$^3$}
\author{Y.~Nanishi$^3$}
\affiliation{$^1$Ioffe Physico-Technical Institute, Russian Academy of
Sciences, 194021 St. Petersburg, Russia}
\affiliation{$^2$LASMEA-UMR 6602 CNRS-UBP, 63177  Cedex, France }
\affiliation{$^3$Ritsumeikan University, 1-1-1 Noji-Higashi, Kusatsu, Shiga 525-8577, Japan}
\begin{abstract}

InN/In nanocomposites with periodical In inclusions amounting  up to 30$\%$ of the total volume exhibit bright emission near 0.7 eV explicitly associated with In clusters. Its energy and intensity depend on the In amount.  The principal absorption edge in the semiconductor host, as  given by a photovoltaic response, is markedly higher than the onset of thermally detected absorption. These findings, being strongly suggestive of plasmon-dominated emission and absorption, are discussed in terms of electromagnetic enhancement taking into account the In parallel-band transitions.
\end{abstract}

\maketitle

Rapidly developed plasmonics has demonstrated advantages of combination of metal and semiconductor to enhance the light emission efficiency \cite{Ozbay}.  This effect results from the interaction of surface plasmons at metallic films or clusters (particles)  with  dipole transitions excited nearby \cite{Moskovits}. The local plasmons at the clusters are effective emitters  with rather high radiative decay rate \cite{Crowell}. It is not trivial to draw the line between the plasmon radiation and  the enhanced emission
due to formation of coupled states.
Most of previous studies considered the clusters situated either on semiconductor surface or inside a dielectric matrix \cite{Ozbay}-\cite{Kreibig}.
We focus on another much less investigated system -- the nanocomposite which contains metallic nano-clusters inside a semiconductor.
This  nanocomposite differs strongly from the semiconductor host. There is no further a compound with a certain band gap, but rather an effective medium whose optical properties are modified by plasmon oscillations.

Such a situation seems to be fully realized in InN with spontaneously formed In clusters \cite{Shubina}. A controversy arose on both the InN optical gap and infrared emission nature  due to the labored discrimination between plasmon-mediated absorption/emission and conventional ones in the case of the active host \cite{Bechstedt}.  It is worth noting that InN was proposed as a promising material for full-spectrum solar cells enough time ago \cite{Wu}. To our knowledge, no data on the photovoltaic effect below 1.5 eV were reported, that indicates likely  unresolved fundamental problems. As there exist certain difficulties in revealing In clusters by transmission electron microscopy (TEM) \cite{Bartel}, we believe that a study of the  InN/In nanocomposites could help to elucidate the metal-related trends.

 \begin{figure} [t]
\includegraphics{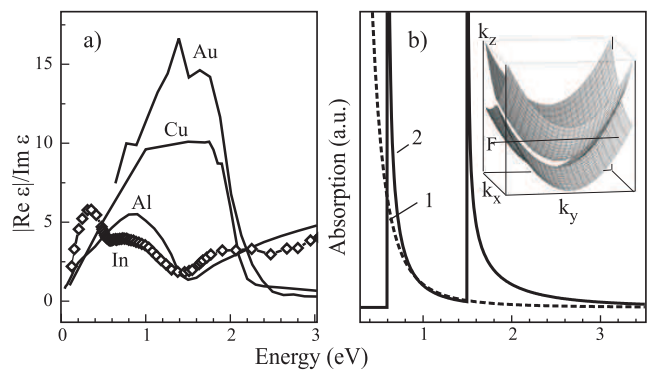}
  \caption{ \label{f1} (a) $|\textmd{Re} \varepsilon|/\textmd{Im} \varepsilon$ spectra shown for In and few other metals \cite{Golovashkin}. (b) Drude (1) and parallel-band (2)  absorption calculated for bulk In. The peaks at 0.6 and 1.5 eV  correspond to (200) and (111) transitions, in notation of \cite{Golovashkin}. The insert presents the parallel bands in \textbf{k}-space with the Fermi energy (F) shown as a guide.
   }
\end{figure}

In the Letter, we present crucial arguments in favor of  plasmonic nature of bright infrared emission and an apparent absorption edge in the InN:In nanocomposites, comprising up to 30$\%$ of excess metal. We show that an optical gap in the InN host is hidden by the plasmonic resonances of the In clusters. In addition, the InN/In nanocomposite is a particular system, where plasmons coexist (and interfere) with the electron transitions between parallel bands,  formed along certain crystal directions in the polyvalent metals (Al, In, \emph{etc.})   \cite{Harrison}. To provide the plasmonic effects, the well-known requirements to the metal permittivity $\varepsilon(\omega)$ at the frequency $\omega$ are: i) $\textmd{Re} \varepsilon (\omega )<0$, and ii) $|\textmd{Re} \varepsilon (\omega )|/\textmd{Im} \varepsilon (\omega )\gg1$. They are  reasonably fulfilled for In in the infrared range, just where the parallel-band transitions take place near the Fermi energy (Fig. 1). Our results show that these transitions suppress selectively the plasmonic excitations.

A set of InN/In structures  was grown by plasma-assisted molecular beam epitaxy (MBE) at the Ioffe Institute by means of periodic deposition of pure In separated by 25 nm of InN.  These insertions are transformed into arrays of cluster agglomerations. The nominal In thickness in different samples was 0 (reference sample), 2, 4, 8, 16 and 48 monolayers (MLs); one ML equals approximately 0.3 nm. Hereafter the thickness in MLs is used for the sample notation. The number of the insertions was 20, except for the 48-ML sample (6 periods). The InN  with the spontaneously formed In clusters, revealed by  TEM  \cite{Bartel}, are also studied (c129). Such clusters occur in the 0-ML sample, despite our efforts to avoid them.
\begin{figure} [t]
\includegraphics{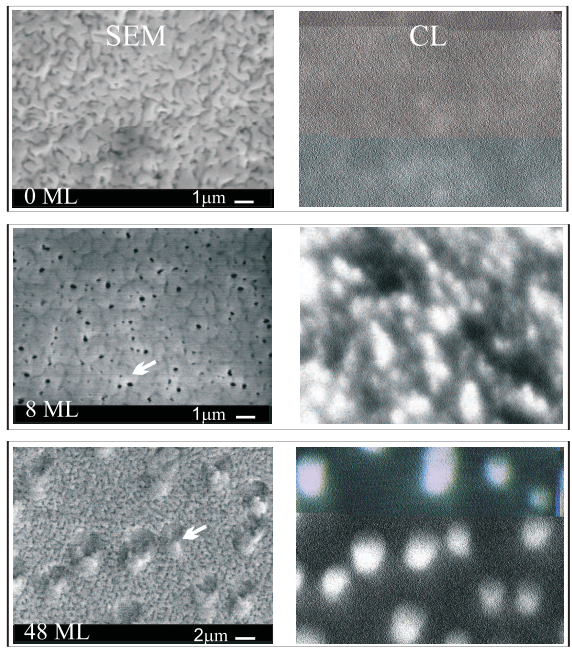}
  \caption{ \label{f2} SEM and CL (150 K) images taken from InN/In samples at 15 keV using a HITACHI S4300SE microscope at the Ritsumeikan University  (the OXFORD CL stage with a 0.6-eV cut-off detector and a filter cutting emission from sapphire). The arrows mark In cluster agglomerations.
   }
\end{figure}

The InN/In composites exhibit emission and absorption edges near 0.7-0.8 eV, i.e. in the range typical of conventional InN films \cite{Davydov}. The striking difference is the spatial correlation of the emission sites with the In clusters, revealed by scanning electron microscopy (SEM) and cathodoluminescence (CL) imaging of the same areas.  The cluster agglomerations are well resolved  in the SEM images as lighter areas due to the heavier In atomic weight. They are almost absent in the 0-ML sample, where the CL is weak and relatively uniform. With the  In amount increase, the bright CL spots appear, coinciding with these agglomerations (Fig. 2).
On an average, the 48-ML sample comprising  1/3 of metal in the periodic structure has the luminescence intensity fivefold higher than the 0-ML film.
The  ``spot" mono-CL measurement shows difference by the factor of $\approx$65 between the intensities at the In agglomerations and the rest area.

The unambiguous spatial correlation of the emission with metallic clusters indicates presumably its relation to the local plasmons.
To verify the plasmonic resonances, we have performed
a comparative study of thermally detected optical absorption (TDOA) and photocurrent (PC). The non-radiative decay of plasmons is accompanied by  transforming their energy into heat. This effect has to be stronger below an InN absorption edge, where it is not suppressed by interband transitions \cite{Bechstedt}. On the contrary, PC is a measure of carrier generation in the spectral range of the interband absorption in a semiconductor. It must reproduce the InN absorption edge.
\begin{figure} [t]
\includegraphics{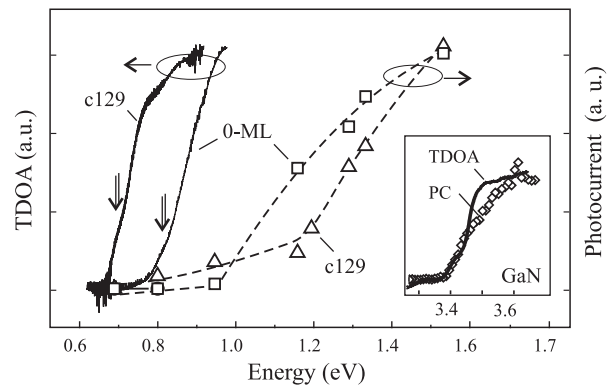}
 \caption{ \label{f1}PC and TDOA recorded in 0-ML and  c129 samples, possessing similar carrier concentration ($\sim2\cdot10^{19}$ cm$^{-3}$). The double arrows denote emission peak energies. The dashed lines are to guide the eye. The inset shows GaN spectra.
   }
\end{figure}

The PC was detected  with the excitation by semiconductor lasers (100 mW power) of various wavelengths, in a planar geometry with  contacts placed on the film surface. The PC spectra (25 K) were recorded  in several samples with a low In amount (Fig. 3). At higher In content, a signal drops down by two orders of magnitude at 1.5 eV; and it cannot be measured below. In accord with the PC data, the principal absorption edge in the host is at 1-1.2 eV. Since the plasmons could enhance the carrier generation, if some host density of states would have been, it must be negligible below this energy.

The TDOA spectra are measured at 0.35 K in a pumped 3He cryostat and then normalized to those of an InAs-bolometer, measured simultaneously. The TDOA edges are found to be  lower in energy  than those of PC (Fig. 3). Previously, it has been demonstrated that an optical gap in non-stoichiometrical InN can be varied due to the strong difference in the N and In atomic orbital energies \cite{Shubina2}. The In gathering into the clusters deplete the InN matrix  by In atoms, shifting its gap  to the higher energy.  Variation of the TDOA edge follows the cluster statistics (see below).
Without plasmons,  the onsets of the PC and TDOA coincide perfectly in a conventional semiconductor, as has been proved by the similar  measurements of GaN (Fig. 3, inset).

The InN/In spectra exhibit other peculiarities which also cannot be explained neglecting plasmons. (i) With the In amount increase, the CL peak shifts gradually  from 0.82 to 0.69 eV, while the principal peak in the TDOA spectra moves to the opposite direction (Fig. 4). (ii) The emission band has frequently two components originating from different areas (Fig. 5). (iii) The TDOA spectra possess reproducibly a dip near 1.5 eV, i.e. at the energy of the (111) parallel-band transitions.

\begin{figure} [t]
\includegraphics{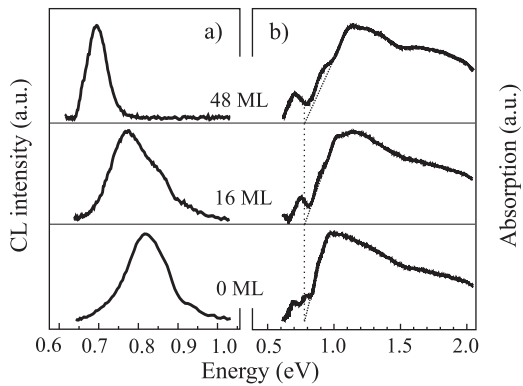}
\caption{ \label{f3} (a) CL  and (b)  TDOA spectra recorded in different samples with carrier concentration (in units of 10$^{19}$ cm$^{-3}$): 0-ML - 2, 16-ML - 4.0, 48-ML - 4.5.  The dotted lines give the TDOA onsets showing a weak concentration dependence.
}
\end{figure}

\begin{figure} [t]
\includegraphics{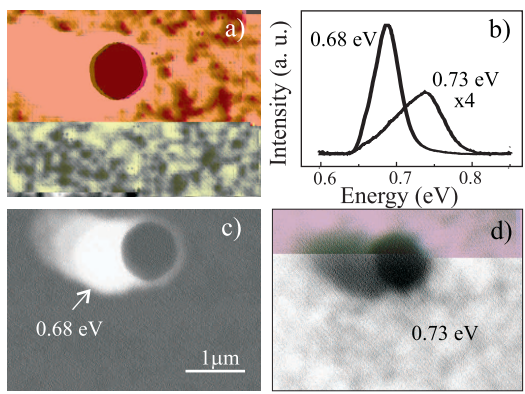}
  \caption{ \label{f2}  Panchromatic CL image (a),  CL spectra measured in the ``spot"-regime (b), and mono-CL images (c, d) taken at energies 0.68  and 0.73 eV, respectively, in a sample around an In droplet (the Ritsumeikan University microscope).
   }
\end{figure}

Plasmonic effects can be considered in terms of the electromagnetic enhancement \cite{Moskovits} implying the local electric field $E$  to be much stronger than the external field  $E_0$, i.e. $|g|=|E/E_{0}|\gg1$. In such a field $E$, both emission and absorption spectra are enhanced as  $|g| ^{2}$.
To analyze the enhancement, we simulate a metal cluster by a spheroid whose rotation axis $c$ is parallel and two other  axes,  $a=b$, are perpendicular to the growth direction.
Surface plasmon polarized along  $i$-th axis of the spheroid provides at its poles the partial enhancement factor
\begin{equation}\label{eq2}
|\tilde{g}_{i}(\omega)| =|\frac{\varepsilon}{\varepsilon_{1}+L_{i}(\varepsilon -\varepsilon _{1})}|.
\end{equation}
Here, $\varepsilon(\omega)$ and $\varepsilon _{1}(\omega)$ denote  the permittivities in  and out of the spheroid, $L_{i}$ is the depolarizing factor.
The factor (1) becomes as large as approximately
${L_{i}^{-1}}|\textmd{Re} \varepsilon|/\textmd{Im}\varepsilon\gg1$
at the resonance condition $\textmd{Re}[\varepsilon_{1}+L_{i}(\varepsilon-\varepsilon_{1})]=0$
which is satisfied by the surface plasmon frequency
$\omega _{i}=\omega _{p}/\left[ \varepsilon _{\infty }+\varepsilon
_{1}\left( L_{i}^{-1}-1\right) \right] ^{1/2}$,
with $\omega_p$  and  $\varepsilon_\infty$ being the plasma frequency and the background dielectric constant of bulk metal.
In general, the local enhancement factor $|{g}_{i}(\omega,\textbf{r})|$  as a function of position $\textbf{r}$  outside the spheroid follows the dipolar field of  $i$-th surface plasmon and varies from (1) at  $i$-th poles down to about
${L_{i}^{-1}}|\textmd{Re} \varepsilon_{1}|/\textmd{Im}\varepsilon\sim1$
at the  pole of an axis normal to the $i$-th axis and in the interior of the spheroid, where the electric field is uniform.
Both $|{g}_{i}|$  and $\omega _{i}$  depend on the aspect ratio $a/c$  via $L_i$; that is why the frequency $\omega _{i}$  can be in the infrared range for surface plasmon polarized along the longest axis with $L_{i}\ll1$.

The enhancement for an In cluster is estimated with the
dielectric function $\varepsilon(\omega)=\epsilon_{\infty}+i4\pi\sigma(\omega)/\omega$,
where the conductivity $\sigma=\sigma_D+\sigma_P$ includes the Drude contribution $\sigma_D(\omega)$ and the term
\begin{equation}\label{eq5}
\sigma_P ( \omega ) =\underset{j}{\sum }\sigma _{j}\frac{\omega
_{j}^{2}}{\omega \sqrt{\omega ^{2}-\omega _{j}^{2}}}\vartheta ( \omega
-\omega _{j})
\end{equation}
describing absorption due to the parallel-band transitions \cite{Harrison}.
The  function $\sigma_D$  is fitted using the data of \cite{Golovashkin}.
In Eq. (2), $\omega _{j}=2U_{j}/\hbar $; $U_{j}$ is the Fourier component
of the lattice pseudopotential; $\vartheta ( \xi ) =0$, if
$\xi <0$, and $\vartheta ( \xi ) =1$, if $\xi >0$.
We take $2U_{( 200) }$=0.6 eV, $2U_{( 111) }$=1.5 eV, $\sigma _{j}=
1\cdot10^{15}$ s$^{-1}$ for both.  For real crystals, to smooth the singular term in Eq. (2) we add to $\omega$  the imaginary part of  $1.6\cdot10^{14}$ s$^{-1}$ equal to the damping parameter in  $\sigma_D$. The  $\varepsilon_1 (\omega)$ function is chosen to match the  $\sim$1-eV optical gap of the host, determined in the PC study.

In composites, the shape and orientation of clusters are random. To get an enhancement averaged over the ensemble, we use a model of spheroids with random ratios $a/c$  and the same volume. Then, the inhomogeneously broadened enhancement spectrum
$G(\omega)=\langle |g(\omega,\textbf{r})|^{2}\rangle$
due to a total set of plasmons is obtained by spatial averaging about a spheroid followed by averaging over the spheroid shapes governed by  $a/c$.
For oblate [prolate] spheroids, their distribution function is non-zero in the interval $1\leq{a/c}\leq(a/c)_{max}$  [$(a/c)_{min}\leq{a/c}\leq1$], where it is presented by a part of Gaussian having the center at $a/c=1$  and a width $\delta$  [$1/\delta$]. Note that $\delta$  is the only fitting parameter in calculating  $G(\omega)$ when the aspect ratio is highly dispersed.
When $\delta$ increases, the maximum in the $G(\omega)$ spectrum undergoes a low-energy shift  owing to additional contributions from spheroids with smaller $L_{i}$, supporting plasmons with smaller  $\omega_i$ [Fig. 6 (a)].
The maximum partial enhancement $|\tilde{g}_{i}(\omega_{i})|^2$ from Eq. (1) may be as high as $10^3-10^4$  near the strongly curved surface of a single spheroid. However, because of a small relative volume of such regions and the dispersion of spheroids over $a/c$, the enhancement $G(\omega)$ is somewhat smaller than $\sim10^2$ at 0.7 eV, in reasonable agreement with the CL intensity contrast.
\begin{figure} [t]
\includegraphics{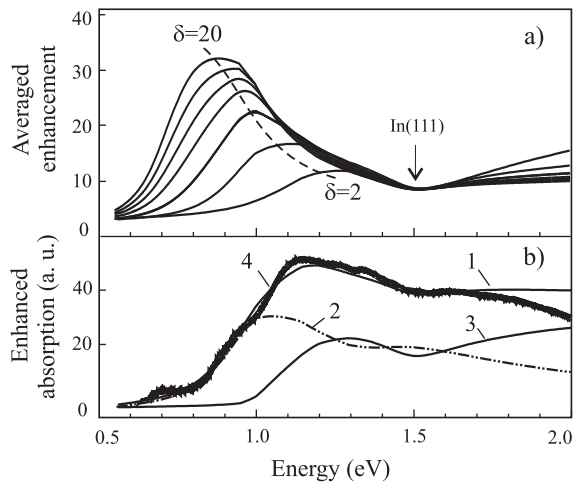}
 \caption{ \label{f1} (a)  Averaged enhancement spectra calculated with  $2<\delta<20$, $(a/c)_{max}=20$ and $\sim$1-eV InN gap. (b)  Enhanced absorption  spectrum (1) and its  constituents from metallic clusters (2) and the InN host (3) calculated  with $\delta=4$. In the TDOA spectrum (4), inessential low-energy interference features are subtracted.
   }
\end{figure}

The  dissipative loss  in $n$-th medium   is  $\sim\omega\textmd{Im}(\varepsilon_n)|{E_n}|^2$ \cite{Landau}. To estimate these losses (heating) in a metal cluster and semiconductor host, the related spectra $G(\omega)$  multiplied by $\textmd{Im} \varepsilon (\omega )$   and  $\textmd{Im} \varepsilon_1 (\omega)$,   respectively, are calculated. With $\delta=4$, this sum fits reasonably the normalized TDOA in the 48-ML sample [Fig. 6 (b)], where the area not influenced by the plasmons is negligible. Below the  host absorption edge, the  signal is exclusively associated with plasmons of the clusters.

Within the plasmon concept, the shift of the CL line and the principal TDOA peak in the opposite directions  (Fig. 4) is not surprising, for the emission and heating are predominantly located in areas whose plasmonic properties are different.
Since the radiative decay rate of plasmons rises strongly  on the characteristic size $a$ of clusters, as $a^{3}$ for spheres \cite{Crowell},
it prevails in the dense cluster agglomerations, where formation of large particles is facilitated.  On the contrary, the plasmon dissipation (heating) dominates in the small clusters uniformly dispersed inside the matrix. With increasing the In amount, homogenization of the cluster shapes,  resulting in the $\delta$ decrease, and the host edge shift in the depleted matrix  contribute to higher energy of the TDOA peak [Fig. 6].

The plasmon frequency can  decrease in the dense cluster agglomerations due to i) shape variation increasing the aspect ratio;
ii) collective interaction between local plasmons \cite{Persson}. We estimated the relative frequency shift of the radiative collective plasmon mode with respect to $\omega_{i}$ of a single cluster to be as high as tens meV. This value is consistent with the energy difference between CL lines originating  from the In-enriched areas adjusting to an In droplet and far from that, where the clusters are well separated  (Fig. 5). When the signal is collected from total area, this effect provides the  doublet emission band.

The calculated absorption spectra have a dip at 1.5 eV, which is similar to that reproducibly observed  in the TDOA.  It appears due to selective suppression of the plasmons by the (111) parallel band transitions, inherent for In.  The (200) transitions, superimposed with the strong Drude absorption, do not exhibit such a peculiarity, but diminish half as much the enhancement nearby. The suppression evidences the enhancement implicitly; without that the  absorption rise might be.

In conclusion, our findings  argue for strong plasmonic contributions to  infrared emission and absorption in InN/In.  We emphasize that it is not valid to compare directly  characteristics of optical processes in the nanocomposites (TDOA,  photocurrent, and emission here). A hidden InN gap is a bright example of that. These results may be important for InN-based devices and for  nanocomposites with clusters of polyvalent metals.

We thank Profs. M.~I.~Dyakonov, A. Kavokin, and Dr. M.~M.~Glasov for fruitful discussions;   Dr. N.~A.~Pikhtin for supplying laser diodes; K.~Kosaka for help in measurements. This work is supported by RFBR, Presidium of RAS, ANR,  and the RFBR-JSPS joint program.

\end{document}